\def\Nat{{\em Nature}}

\def\spose#1{\hbox to 0pt{#1\hss}}
\def\approxlt{\mathrel{\spose{\lower 3pt\hbox{$\sim$}}
        \raise 2.0pt\hbox{$<$}}}
\def\approxgt{\mathrel{\spose{\lower 3pt\hbox{$\sim$}}
        \raise 2.0pt\hbox{$>$}}}

\def\multleft#1{\hbox to size{\vbox {\halign {\lft{##}\cr #1}}\hfill}\par}
\def\multright#1{\hbox to size{\vbox {\halign {\rt{##}\cr #1}}\hfill}\par}

\def\boxit#1{\vbox{\hrule\hbox{\vrule\kern3pt\vbox{\kern3pt
          #1 \kern3pt}\kern3pt\vrule}\hrule}}

\def\cm{{\rm\thinspace cm}}

\def\erg{{\rm\thinspace erg}}

\def\s{{\rm\thinspace s}}

\def\pcmsq{\hbox{$\cm^{-2},$}}

\def\ergpcmsqps{\hbox{$\erg\cm^{-2}\s^{-1}\,$}}
\def\ergpcmsq{\hbox{$\erg\cm^{-2}\,$}}

\def\pcmsq{\hbox{$\cm^{-2}\,$}}

\documentclass{aastex}
\usepackage{emulateapj5}
\usepackage{psfig}

\input epsf           
\makeatletter

\newenvironment{inlinefigure}{%
\def\@captype{figure}%
\noindent\begin{minipage}{0.999\linewidth}\begin{center}}
{\end{center}\end{minipage}\smallskip}
\makeatother

\slugcomment{Submitted to the Astrophysical Journal Letters}

\begin{document}

\title{Imaging Large Scale Structure in the X-ray Sky }

\author{Y. Yang\altaffilmark{1,2}, R. F. Mushotzky\altaffilmark{2}, A. J. Barger\altaffilmark{3,4,5}, L. L. Cowie\altaffilmark{5}, D. B. Sanders\altaffilmark{5,6}, A. T. Steffen\altaffilmark{3}}

\altaffiltext{1}{Department of Astronomy, University of Maryland, College Park 20742-2421}
\altaffiltext{2}{Laboratory for High Energy Astrophysics, Goddard Space Flight Center, Code 660, NASA, Greenbelt, MD, 20770}
\altaffiltext{3}{Department of Astronomy, University of Wisconsin at Madison, 5534 Sterling Hall, 475, Madison, WI 53760} 
\altaffiltext{4}{Department of Physics and Astronomy, University of Hawaii, 2505 Correa Road, Honolulu, HI 96822}
\altaffiltext{5}{Institute for Astronomy, University of Hawaii, 2680 Woodlawn drive, Honolulu, HI 96822} 
\altaffiltext{6}{Max-Planck-Institut fur Extraterrestrische Physik, D-85740, Garching, Germany}

\begin{abstract}
We present the first results from a wide solid angle, moderately 
deep {\it Chandra} survey of the Lockman Hole North-West region. 
Our 9 ACIS-I fields cover an effective solid angle of 0.4 deg$^{2}$ and 
reach a depth of  $3 \times 10^{-16}$~\ergpcmsqps in the 0.4--2 keV band and 
$3 \times 10^{-15}$~\ergpcmsqps in the 2--8 keV band. The best fit logN-logS
for the entire field, the largest contiguous {\it Chandra} field yet observed, 
matches well onto that of the  {\it Chandra} Deep Field North. We show that 
the full range of the `cosmic variance' previously seen in different 
{\it Chandra} fields is reproduced in this small region of the sky. 
Counts-in-cells analysis shows that the hard band sources are
more strongly correlated than the soft band sources.        
\end{abstract}


\section{Introduction}
Recent {\it Chandra} and {\it XMM} observations have resolved over 85\% of the 
2--8~keV X-ray background (XRB) into discrete sources, presumably active 
galaxtic nuclei (AGNs; Mushotzky et al. 2000; Brandt et al. 2001; 
Tozzi et al. 2001; Campana et al. 2001; Cowie et al. 2002; Giacconi et al. 
2002; Hasinger et al. 2001).  However, the nature of the sources is unclear, 
with many of  the faint sources showing little or no optical activity (Barger 
et al. 2001, 2002; Hornschemeier et al. 2001; Rosati et al. 2002).  There is  
a large field-to-field variance in the source counts in the {\it Chandra} 
deep surveys in the  2--8~keV band (Cowie et al. 2002).
This cosmic variance is generally believed to arise from the 
underlying large-scale structure (LSS) that 
is traced, in some fashion, by the {\it Chandra} sources. 
Clustering of the XRB sources in redshift space has been seen in both 
{\it Chandra}  Deep Fields (Barger et al. 2002; Hasinger 2002),
 indicating that LSS does exist in the XRB source distribution.  
From a cosmological point of view, AGNs are expected to be highly 
biased tracers of cosmic structure formation at medium to large redshifts.
The XRB sources, which have $\sim10$ times the areal density of 
optically selected  AGNs, provide for the first time a sufficiently 
high density to be used as tracers of the LSS.

\begin{figure*}[t]
\centerline{\psfig{figure=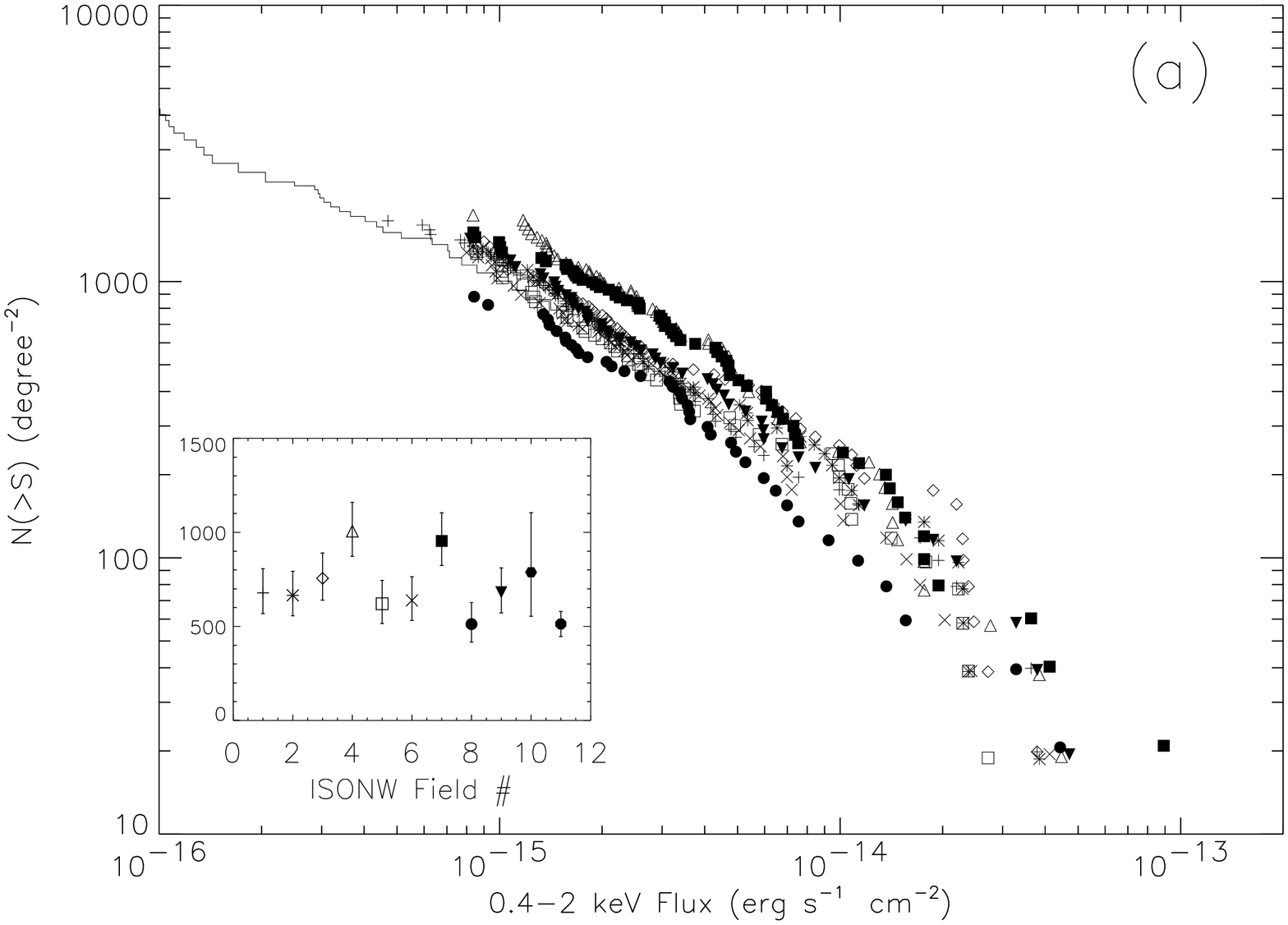,width=0.4\textwidth,angle=90}\psfig{figure=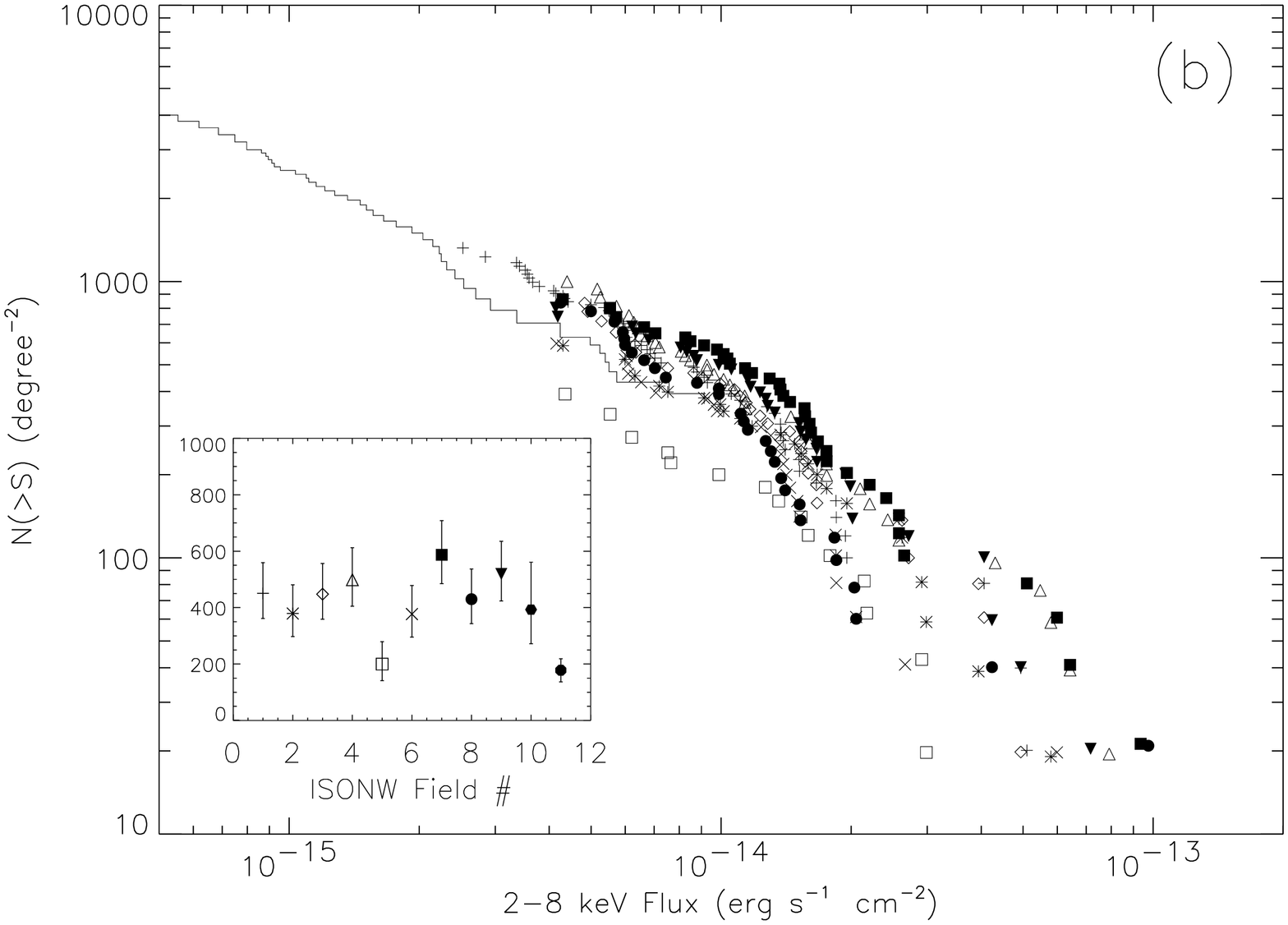,width=0.4\textwidth,angle=90} }
\centerline{\psfig{figure=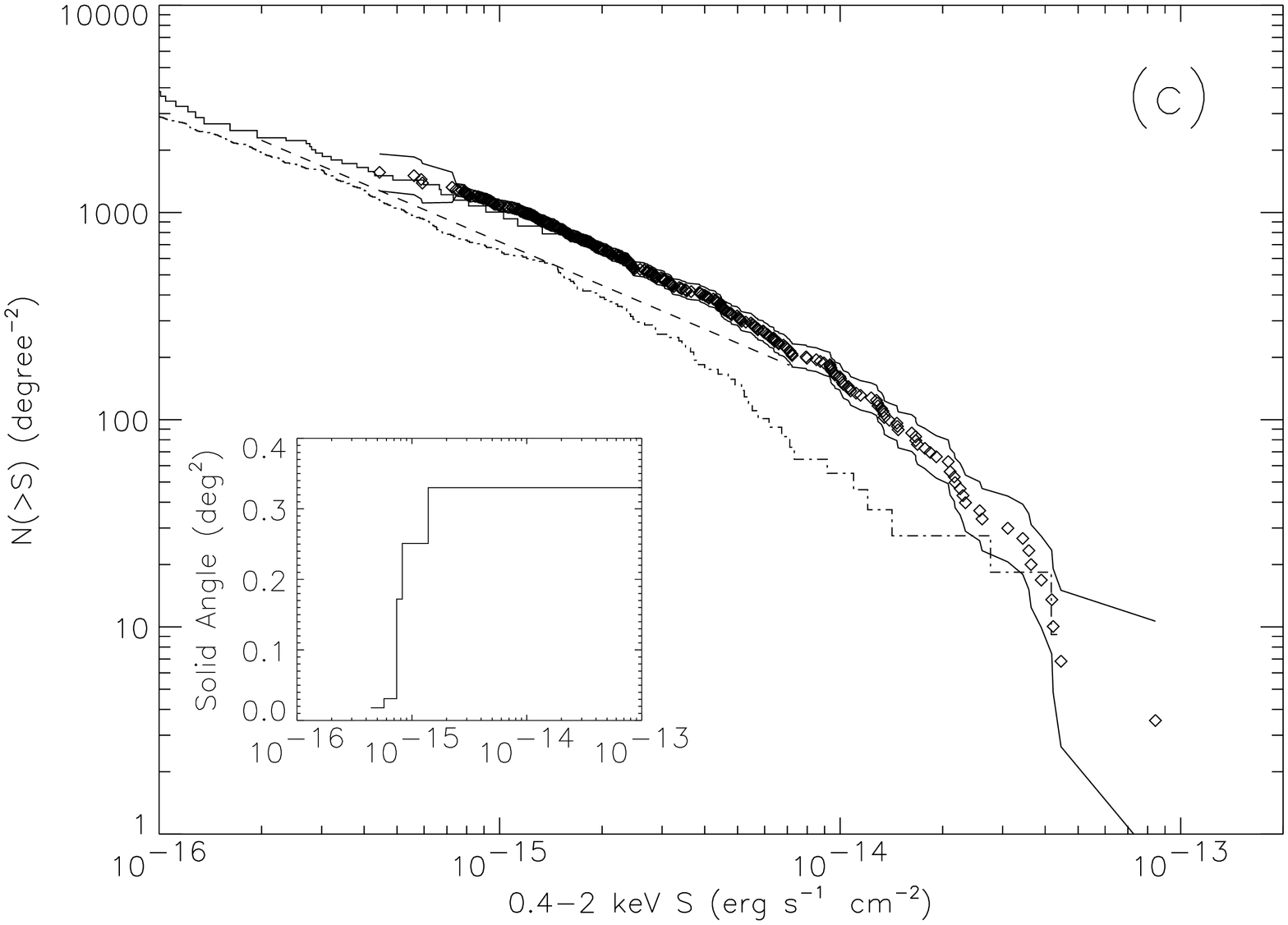,width=0.4\textwidth,angle=90}\psfig{figure=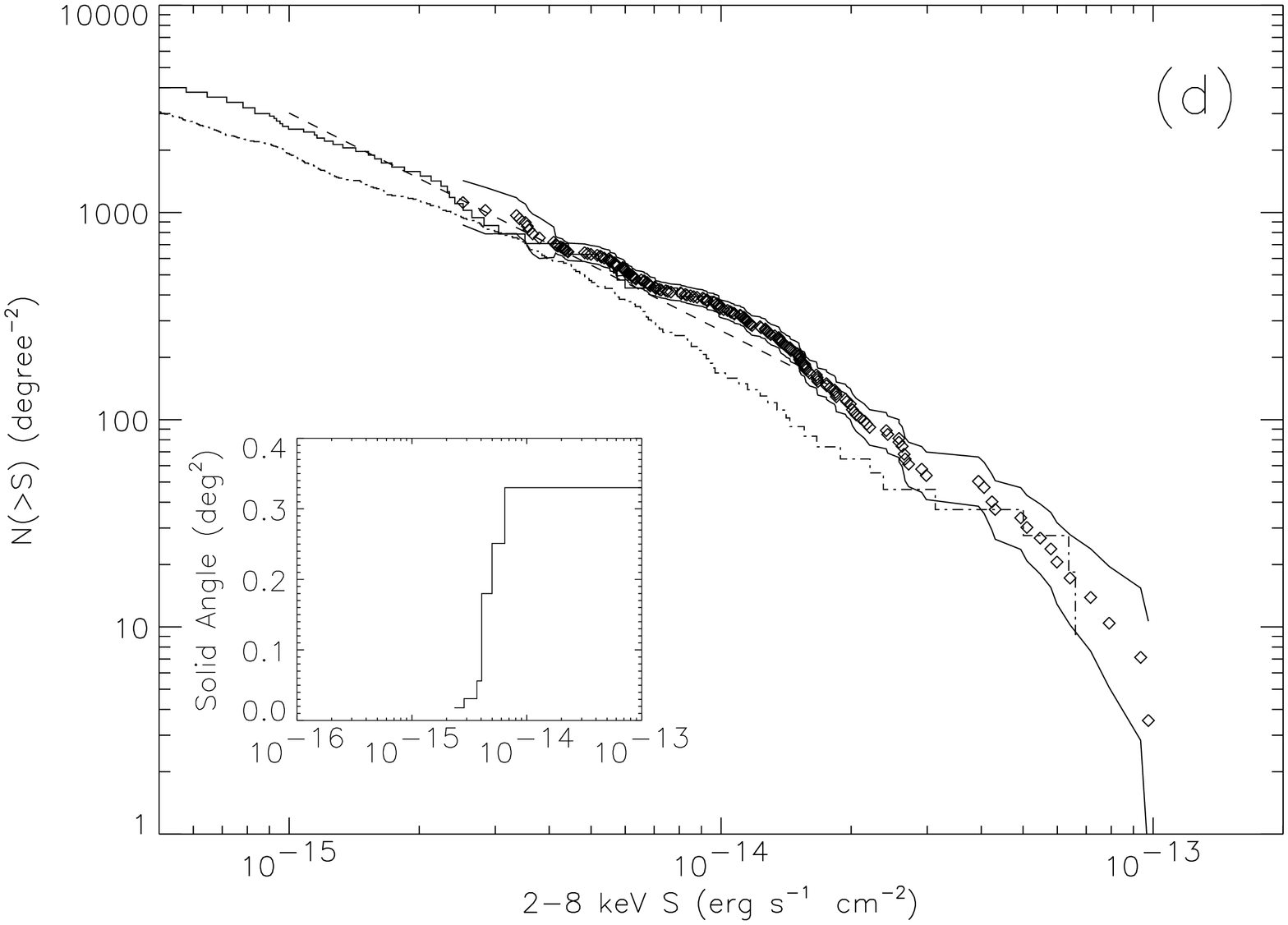,width=0.4\textwidth,angle=90} }
\caption{The cumulative number counts in the 9 fields (a), 
(b) and the whole field (c), (d).
For each (a) and (b), the insert shows the number counts of each 
ISONW field at flux level of $2\times 10^{-15}$~\ergpcmsqps (soft) 
and  $9 \times 10^{-15}$~\ergpcmsqps (hard). The solid line is the 
logN-logS of CDF-N. Field numbers 10 and 11 represents 
the CDF-N and CDF-S value respectively. 
The highest counts in the 9 fields are about 2.3 (soft) and 3.3 (hard) 
times higher than that from CDF-S, which has the lowest normalization among 
all the deep surveys (Cowie et al. 2002). For the hard band, number counts of 8 of the 9 fields are $\sim 1\sigma$ about the mean, but ISONW5 is much lower. For each (c) and (d), 
cumulative number counts for the whole field (diamonds with solid 
lines representing the  Poisson error) is compared with that from CDF-N 
(solid histogram), CDF-S (dash-dotted histogram) and 
SSA13 (dashed line). The inserts shows the sensitive sky area vs flux. }
\end{figure*}

The clustering at various scales of X-ray selected AGNs has been pursued 
previously using observations from {\it HEAO-1} ( e.g. Barcons \& Fabian 1988;
 Mushotzky \& Jahoda 1992) and {\it ROSAT} (e.g. Soltan et al. 1997; 
Vihklinin \& Forman 1995). 
The {\it ROSAT} results based on deep pointings (Carrera et al. 1998) show a 
much smaller amplitude of correlation than predicted from optical samples, 
while an analysis of the {\it ROSAT} North Ecliptic Pole sample, which is 
shallower but covers a wider solid angle (Mullis 2002), comes to the 
opposite conclusion.  
It has been claimed that at a flux level below $1.5 \times 10^{-15} $ 
\ergpcmsqps (0.5--2 keV) the XRB exhibits no clustering (Soltan \& 
Hasinger 1994). This contradicts the observed cosmic variance in the
ultradeep surveys by {\it Chandra} and {\it XMM}. How to combine these 
seemingly disparate results is one of the new mysteries of the XRB.

To allow a direct measurement of the auto-correlation functions (ACFs) of 
the XRB sources, and to optically identify these 
objects to obtain their redhshifts, one needs arcsecond spatial resolution, 
a wide contiguous field of view, and sufficient sensitivity. 
To achieve these goals we have performed a large solid angle, 
moderately deep {\it Chandra} survey of the Lockman Hole North-West region.
This survey currently has a contiguous sky coverage of 
$\sim 0.4$ deg$^{2}$ and is sensitive to X-ray flux levels of  
$3 \times 10^{-16}$ \ergpcmsqps (0.4--2 keV) 
and $3 \times 10^{-15}$ \ergpcmsqps (2--8 keV). 
Most of the XRB is resolved at these flux levels (Cowie et al. 2002).   
 
\section{Observations and Data Reduction}   
The survey covers the Lockman Hole North-West region centered 
at $\alpha=10^{h}34^{m}$, $\delta=57\arcdeg40\arcmin$ (J2000). 
The region has very low Galactic column density ($N_{H} \equiv 
5.72\times 10^{19}$\pcmsq, Dickey \& Lockman 1990) and was covered 
by the deepest 170$\mu$m ISOPHOT field observed from {\it ISO} 
(hereafter ISONW).
The field  has also been observed at 850$\mu$m with the SCUBA camera on the 
James Clerk Maxwell Telescope and at cm wavelengths with the VLA, as well as 
at optical and near-IR wavelengths with the Subaru (using the unique 
Suprime-Cam instrument) and Keck telescopes. The ISONW region will be intensively observed 
with {\it SIRTF} as part of the LEGACY program. The multiwavelength analysis
of this field will be presented in subsequent papers.

The observation consists of 9 ACIS-I pointings (labeled as ISONW1--ISONW9) 
separated from each other by $\sim 10\arcmin$ to allow close to uniform 
sky coverage.
ISONW1 has an exposure time of $\sim 70$~ks while the other 8 pointings have 
exposure times of $\sim 40$~ks. 

The data were reduced with CIAO 2.2.1 and CALDB 2.15. The CXC 
{\it Science Threads} were followed in data preparation. 
The resulting event lists were binned into 
2 energy bands, the soft (0.4--2 keV) and the hard (2--8 keV). 

Point sources were detected for each pointing in both energy bands 
with {\it wavdetect} within the CIAO package. The wavelet scales 
of square root series $1,\sqrt 2, 2, 2 \sqrt 2, 4, 4 \sqrt 2,8$ and 
a false  detection threshold of $1 \times 10^{-7}$  were used.   
Spectrally weighted  monochromatic exposure maps were created, 
assuming a power law with photon index of 1.2 for
the hard band and 1.4 for the soft band (Mushotzky et al. 2000; Barger et al. 
2001). Count rates were converted to flux assuming the above   
power law spectra with only Galactic absorption. The conversion factors  
are $4.74 \times 10^{-12}$ \ergpcmsq (soft) and 
$2.34 \times 10^{-11}$ \ergpcmsq (hard).
The degradation of the ACIS low energy quantum efficency during 
the flight was corrected using the measurements of the ACIS team. 

The catalogs for all observations were merged. Source properties for 
objects detected in more than one observation (due to the overlapping
of fields) were taken from the field in which the source had the smallest
off-axis angle. In the soft band, 431 sources were detected, and in the hard 
band, 278. The combined catalog contains 554 sources.

\section{Analysis and Results}  
Since the effective area decreases and the point spread function (PSF)  
increases with off-axis angle, the sensitivity  is not 
uniform across the field. We quantified this with Monte-Carlo simulations. 
First we constructed background maps by removing the wavelet detected 
sources from the observed images and filled the holes with Poisson noise 
sampled from regions surrounding the sources. Sources with fluxes drawn from 
the LogN-LogS derived from the {\it Chandra} deep fields (Cowie et al. 2002;
 Garmire 2002) were generated and distributed uniformly within $8\arcmin$ from
 the aim point. PSF images of each source were made using {\it mkpsf}.  
The exposure maps described in the previous section were then applied to 
simulate the effect of vignetting. The simulated sources were added to 
the background maps to create simulated images. About 100 simulations were 
performed for each band and exposure (for details see  Yang et al. 2003). 
Above flux thresholds of 10 cts/exposure (soft) and 12 cts/exposure (hard), 
the detection is complete and the derived fluxes are consistent with the 
input values. We use the flux unit in cts/exposure 
because we found the {\it wavdetect} exposure-corrected count 
rates threshold in such unit only depends weakly on the exposure time in 
our observations due to the low backgrounds. We obtain the same results 
by assuming a uniform background and  the {\it wavdetect} detection thresholds 
described in Freeman et al.(2002).  Within $4.5\arcmin$ of the aim point, 
the detection is complete above 5 cts/exposure in the soft band and 
7 cts/exposure in the hard band. 

Above flux thresholds of $1.2 \times 10^{-15}$ \ergpcmsqps (soft) and
$8 \times 10^{-15}$ \ergpcmsqps (hard), the source catalog
in the combined field within $8\arcmin$ from each aim point is complete. 
This defines a flux limited sample that contains 115 hard band and 298 
soft band sources. The solid angle covered by the complete sample field is 
0.33 deg$^{2}$.  

We constructed the cumulative number counts (LogN-LogS)
for each of 9 fields, as well as the whole merged catalog 
(Fig. 1). The Eddington bias found at these flux levels from 
the simulations is small and can be ignored.

%
%
\begin{inlinefigure}
\centerline{\psfig{figure=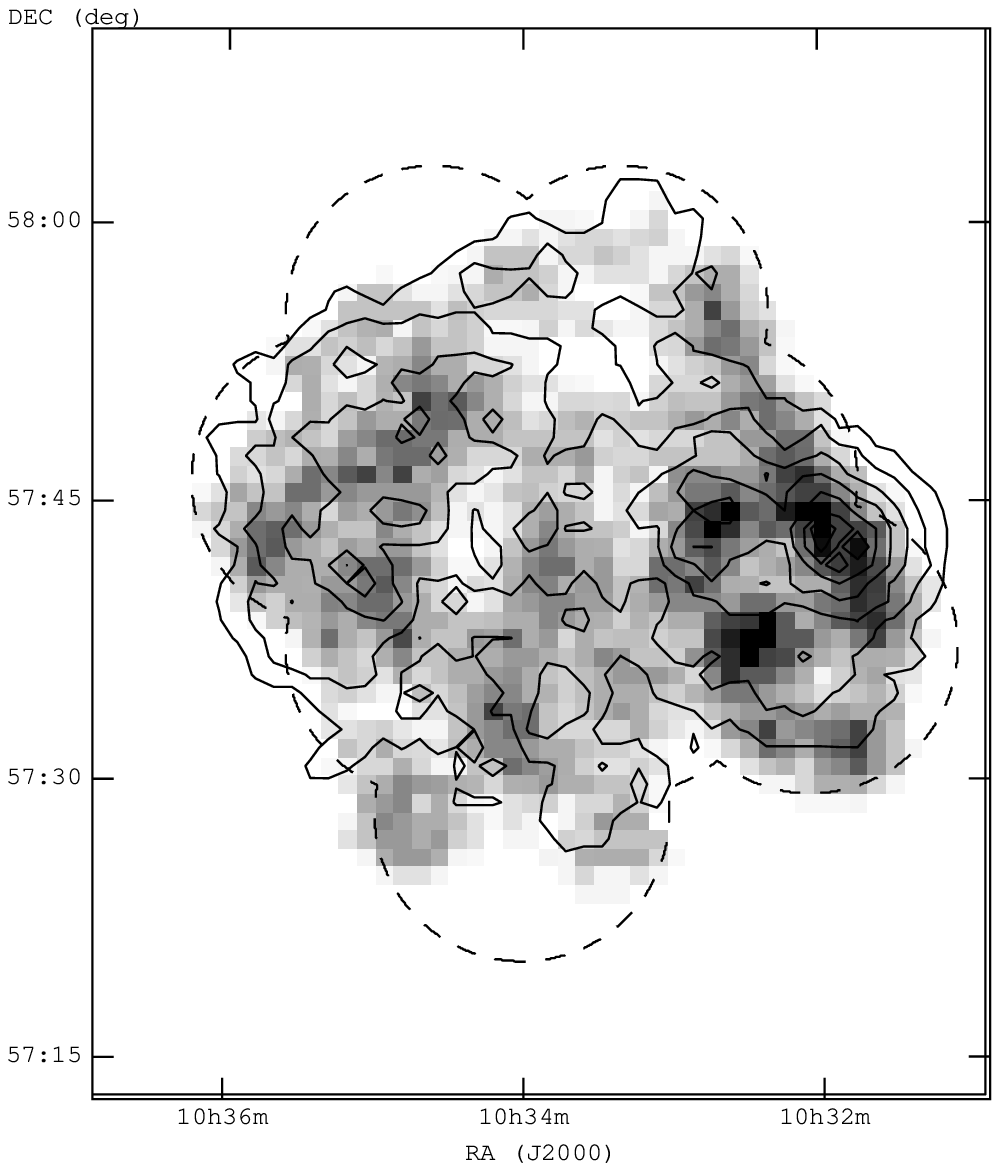,width=0.9\textwidth,angle=0}}
\caption{
The density map of the flux limited samples (region outlined with a dashed line). The maps have been adaptively smoothed using {\it fadapt} in {\it FTOOLS} with a 6 counts kernel counts threshold. The gray scale map represents the source density in the soft band. The contours shows the source density in the hard band. The 7 contour levels are proportional to the logarithm of the source density. }
\label{fig2}
\addtolength{\baselineskip}{10pt}
\end{inlinefigure}

Fig. 1a and 1b show comparisons of the LogN-LogS in the 9 pointings. 
Although the overlapping of fields
could minimize the differences between them, this is the simplest  way to 
demonstrate the cosmic variance because exactly the same procedures 
were used in each field and 8 of the 9 fields have virtually identical 
exposures. Variance is seen in both the soft and hard bands.
The full range of variance seen in the {\it Chandra}
deep surveys published to date is reproduced in this set of 
contiguous fields.

The LogN-LogS of the whole field is compared with that from {\it Chandra}
deep field North (CDF-N; Brandt et al 2001), the {\it Chandra} Deep Field
south (CDF-S, Rosati et al. 2001) and SSA13 (Mushotzky et al. 2000) in 
Fig.~1c and 1d. 
Our LogN-LogS connects smoothly with that of CDF-N, which has the highest 
normalization of all the published {\it Chandra} deep fields. 

The LogN-LogS distribution of the combined fields is modeled with power 
laws in the form of $N(>S)=A(S/S_{0})^{-\alpha}$, using the area 
weighted maximum likelihood method (Murdoch, Crawford \& Jauncey 1973).
For the soft band,  between $2 \times 10^{-15}$ and $10^{-14}$ 
\ergpcmsqps, $A=630,S_{0}=2\times10^{-15}$, and $\alpha=0.72\pm0.18$; 
above $1 \times 10^{-14}$ \ergpcmsqps, $A=152,S_{0}=1\times10^{-14},
and \alpha=1.58\pm0.23$. For the hard band above 
$1 \times 10^{-14}$ \ergpcmsqps, the best fit parameters are 
$A=359,S_{0}=1\times10^{-14}, and \alpha=1.56\pm0.16$.

To visualize the source distributions, we used adaptive smoothing
to create density maps of the flux limited sample (Fig.2). 
Structure is visible in both the soft and hard band maps, but
the hard band sources are more clustered.  
To test whether the observed over-density could arise from 
Poisson fluctuations, we employed the likelihood test described in 
Carrera et al.(1998). Comparing with 10000 simulated samples which are
Poisson  distributed, we found 97.67\% of the soft and 99.99\% of the  
hard band simulations had better likelihood than the observations.  
The significance of clustering is therefore
2$\sigma$ (soft) and 4$\sigma$ (hard).

Using the source distribution in cells tiling the field (counts-in-cells),
we can estimate the correlation scales of the sources.
The variance of counts-in-cells defined as
$\mu_{2} \equiv \langle(N-\bar{N})^{2}\rangle$, 
where $\bar{N}$ is the mean counts in the cell, is directly related to the 
angular correlation function (Peebles 1980) by
\begin{equation}
\mu_{2}=\bar{N}+\frac{\bar{N}^{2}}{\Omega^{2}}\int{w(\theta)d\Omega_{1}d\Omega_{2}}
\end{equation} 
where $\Omega$ is the cell size. The first term is the Poisson fluctuation. 
For correlation functions with a power-law form $ w(\theta)=({\theta}/{\theta_{0}})^{1-\gamma}$ (where $\gamma$ is the power law index of the spatial 
correlation function) and square cells with size 
$\Omega=\Theta \times \Theta$~deg$^{2}$, the integration can be obtained as
(Totsuji \& Kihara 1969; Lahav \& Saslaw, 1992), 
\begin{equation}
\sigma^{2} \equiv \frac{\mu_{2}-\bar{N}}{(\bar{N}/\Omega)^2}=C_{\gamma}\theta_{0}^{1-\gamma}\Theta^{5-\gamma}
\end{equation}
where $\sigma^{2}$ is the normalized variance. $C_{\gamma}$ is a function of $\gamma$ and is calculated numerically. We calculate $\sigma^{2}$ for square
cells of various sizes that tile the whole field. 
By fitting the $\sigma^{2}-\Theta$ relation we should be able to estimate 
$\theta_{0}$ and $\gamma$. We found the present data cannot  
constrain both parameters accurately. By fixing $\gamma=1.8$, the 
``universal'' slope measured in galaxies and in groups and clusters of 
galaxies (Bahcall 1988), and 
minimizing $\chi^2$, we found $\theta_{0}=40 \pm 11\arcsec $ and $\theta_{0}=4 \pm 2 \arcsec$ for the hard and soft band sources respectively. 
While the soft band sources agree very well with the angular correlation scale
previously seen in {\it ROSAT} surveys (Vikhlinin \& Forman, 1995), the hard band sources are much more strongly correlated. 

The striking difference in clustering between the soft and hard band sources 
indicates the hard sources which are not detected in the soft band are highly 
clustered. About 60\% of the hard-band-only sources lie in overdense regions
which form a `band' connecting the `lumps' on the western and eastern side of 
the field (Fig. 2). This band includes only  about 1/3 of solid angle of the 
whole field. The counts-in-cells analysis (Fig. 3) also indicates that these 
hard-band-only sources have larger correlation scales than the soft 
band or hard band sources.

\section{Discussion} 
{\it ASCA} observations have shown that the rms variance of the 2--10 keV 
XRB on a scale of 0.5~deg$^{2}$ is $\sim 6\%$(Kushino et al. 2002). With a sky 
coverage of 0.4~deg$^{2}$ and a depth that resolves  
$> 50\%$ of the hard band XRB, the normalization of LogN-LogS derived from 
our observations should be very close to the ``true'' value. 
The fact that the observed LogN-logS connects onto the CDF-N field at low 
fluxes then indicates that $>90\%$ of the XRB 
is resolved using the {\it ASCA/ROSAT} XRB normalization (Chen et al. 1997). 
The main uncertainty is the normalization of XRB itself.  
 
The LSS seen in our field has reproduced the cosmic
variance observed previously in deep field surveys.  
It is noticeable (Fig. 1b) that on scales of a 
{\it Chandra} field the variance is demonstrated as holes rather than lumps,
i.e., the number counts in 9 fields are close to the mean, while only 2
field have very low value (ISONW5 \& CDF-S).   
This indicates the existence of voids in the X-ray LSS, which should, 
within a factor of a few, be of the same angular size as an ACIS-I field. 
Most of the hard-band-only sources cluster in relatively small regions, 
which may be topologically connected. If this is not a result of a 
projection effect, then it is the 
first time a wall-like structure has been seen in the X-ray sky.     

We have found more variance in the hard band than in the soft band, 
consistent with the lower mean redshift of the hard X-ray sources.
However, the difference in the mean redshift cannot account for most of 
the large difference in the soft and hard band angular correlation length
found by the counts-in-cells statistic. The mean redshift found previously
for the {\it Chandra/XMM} deep fields sources is $\sim 0.8$.  
Unless most of the hard band sources are at very low redshift, which does 
not seem likely because we see no low redshift spikes in the redshift 
distribution of deep field surveys, the relatively low redshifts for the 
hard X-ray selected sources could not account for the factor of 10
larger differences in correlation scales seen in these sources. 
The similarity of the redshift distributions for the sources found in 
CDF-N and in an observation with a similar exposure to ours 
(Castander et al. 2003), indicates that the luminosity functions have 
likely been sampled below the ``knee'' for all redshifts
in our observations. Thus we feel that the stronger correlation function 
for the hard sources seen in our data is likely due to their stronger 
spatial correlation rather than a redshift effect.

The large variance in the {\it Chandra} source counts 
indicates that they must be highly biased tracers of matter. 
Direct calculation of the bias will require redshifts and an understanding 
of how the luminosity function changes with redshift, but it is already clear 
that the {\it Chandra} sources show much more variance than galaxy counts 
(Cohen et al 2000) at similar optical magnitudes. 

\begin{inlinefigure}
\centerline{\psfig{figure=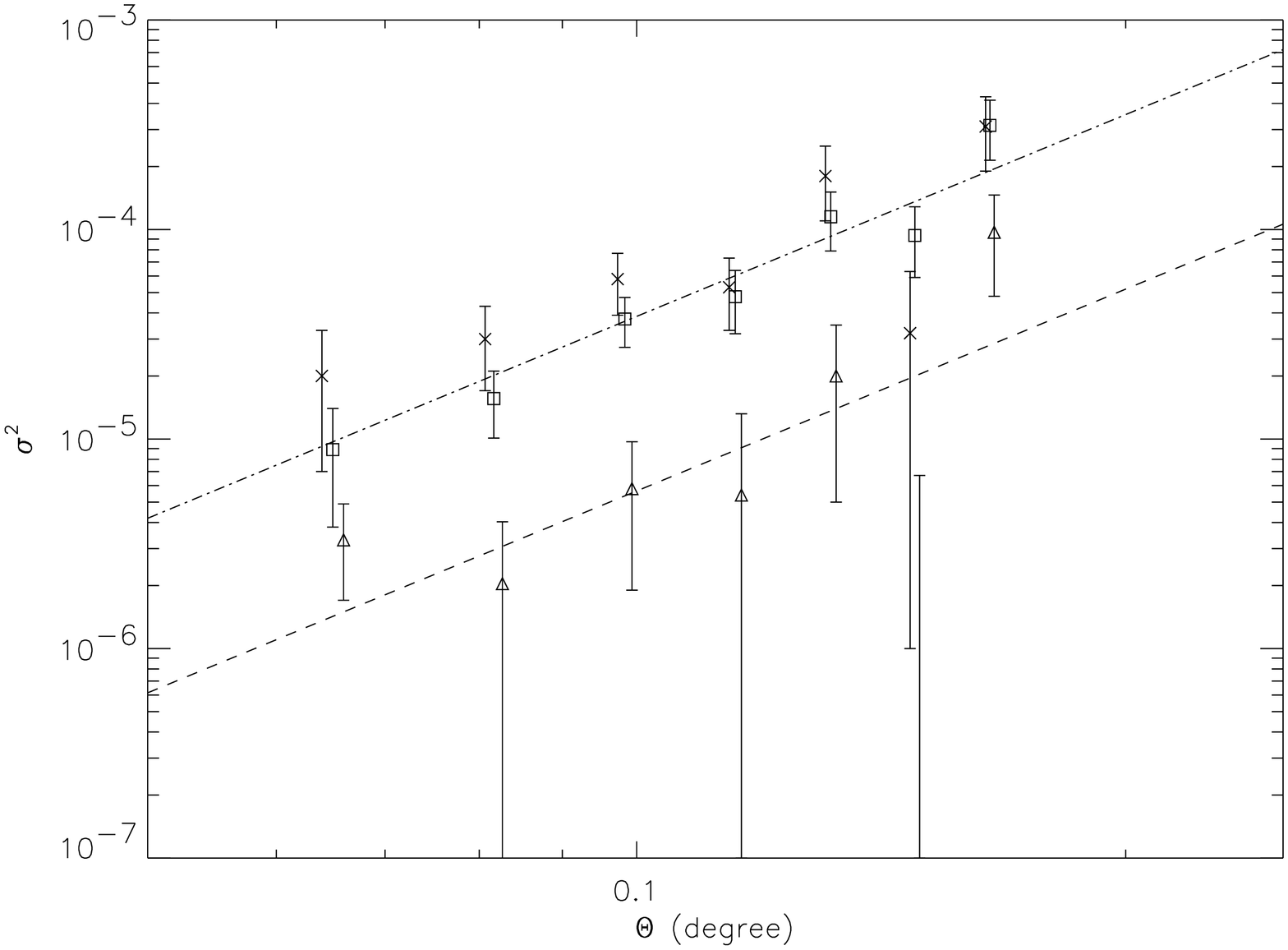,width=0.9\textwidth,angle=90}}
\caption{
Normalized variance $\sigma^{2}$ measured with different sized square cells which tiles the whole field. Squares: the hard band sources; triangles: the soft band sources. Crosses: the hard band only sources.  The errors are estimated via a  boots-trap technique. The lines shows the best fit variance with $\gamma=1.8$(fixed) and  $\theta_{0}=40\arcsec \pm 11$ for the hard band, $\theta_{0}= 4\arcsec \pm 2$ for the soft band sources. The hard-band-only points are not fitted due to the small number of sources. }
\label{fig3}
\addtolength{\baselineskip}{10pt}
\end{inlinefigure}

\acknowledgements

We thank CXC for their excellent work, 
G. Garmire and N. Brandt for providing the CDF-N logN-logS, 
C. Reynolds, K. Arnold, K. Jahoda, D. Davis, K. Kuntz, 
D. Smith and A. Young for very helpful discussions. 
The project is funded under IDS program of 
R. Mushotzky, NSF grants AST-0084847 (A.J.B.) and 
AST-0084816 (L. L. C. ), the University of Wisconsin
Research Committee with funds granted by the Wisconsin Alumni
Research Foundation (A.J.B.) and the Alfred P. Sloan Foundation (A.J.B.). 
D. B. S. gratefully acknowledges the hospitality of the 
MPE and is grateful for support from a senior award from the 
Alexander von Humboldt-Stiftung and from NASA through Chandra Award 
number GO2-3191C.


\end{document}